\documentclass{article}
\usepackage{graphics}
\usepackage{psfig}
\begin{document}

\title{Critical dynamics of DNA denaturation
\footnote{ \uppercase{P}roceedings of the conference
on {\em
`` \uppercase{L}OCALIZATION AND 
ENERGY TRANSFER IN NONLINEAR SYSTEMS"},
\uppercase{J}une 17-21, 2002, \uppercase{S}an 
\uppercase{L}orenzo de \uppercase{E}l \uppercase{E}scorial, \uppercase{M}adrid, 
\uppercase{S}pain. \uppercase{T}o be published by \uppercase{W}orld 
\uppercase{S}cientific.}
\footnote  {\uppercase {W}ork partially
supported by 
\uppercase {EU} contract \uppercase
{HPRN-CT-1999-00163 (LOCNET} network).    }
} 
\author{N. Theodorakopoulos  \\
\it Theoretical and Physical Chemistry Institute,\\
\it National Hellenic Research Foundation,\\
\it Vasileos Constantinou 48, 116 35 Athens, Greece \\
\\
M. Peyrard and T. Dauxois\\
\it Laboratoire de Physique, UMR-CNRS 5672, \\
\it ENS Lyon, 46 All\'{e}e d'Italie, \\
\it 69364 Lyon C\'{e}dex 07, France}
\maketitle

{\small
We present detailed molecular dynamics results for the displacement
autocorrelation
spectra of the Peyrard-Bishop model of thermal DNA denaturation.
As the phase transition is approached, the spectra depend
on whether the wavelength is smaller than, or exceeds the correlation
length. In the first case, the spectra are dominated by a single
peak, whose frequency approaches the bare acoustic frequency
of the harmonic chain, and whose linewidth approaches zero
as $T_c-T$. In the second case, a central peak (CP) feature is dominant,
accounting for most of the weight; the linewidth of the CP appears
to be temperature-independent. We also present force autocorrelation
spectra which may be relevant for analyzing the statistical
properties of localized modes.
}

\section{Introduction}

The thermal denaturation of DNA, i.e. the separation
of the two strands upon heating, is a typical thermodynamic
instability.  It can be modelled along the lines of other
thermodynamic instabilities (e.g. wetting, solid-on-solid adsorption),
by associating a single, one-dimensional  coordinate with the
distance of a base pair\cite{PBI}. Details can be found in this
volume\cite{nthrev} and in the original literature cited there.

The equilibrium properties of the system near the phase transition
are characterized by a divergent correlation length $\xi$, and a
discontinuity in the specific heat; in other words, this is a
second-order transition; the feature which sets it apart from
other structural, or order-disorder transitions is that, as the
transition temperature is approached from below, the order
parameter diverges, i.e. the low-temperature phase becomes
continuously unstable.

In this paper, we present results for the dynamical correlations
of the order parameter, obtained by numerical simulation.
At low and intermediate temperatures,
the spectra appear to be dominated by the properties of
localized anharmonic motion (``discrete breathers'').  As
the critical temperature is approached from below,
the spectra depend solely on whether the
wavelength is smaller or larger than the correlation length. In
the first case, they reflect the dynamics of ``islands'' of the
high temperature phase. In the second case, they are
dominated by a strong central peak, whose width appears
to depend on the wavevector, but not on the temperature.

The paper is structured as follows: Section 2 introduces the
notation and numerical procedure. Section 3 presents the
main results, and an analysis along the lines of
relaxational/oscillational phenomenology. Section 4 is a
sketch of a tentative, alternative theory, along the lines
of the Mori-Zwanzig projection operator formalism.
Section 5 presents a brief summary and discussion.


\section{Notation and  numerical procedure}
We consider the ``minimal" Hamiltonian model of homogeneous DNA denaturation
proposed by Peyrard and Bishop\cite{PBI}(PB),
\begin{equation}
H(y)= \sum_{n}\left[ \frac{ p_{n}^{2} } {2}
+\frac{1}{2R} ( y_{n}-y_{n-1})^{2}
+V(y_{n})
\right] \quad,
\label{eq:trham}
\end{equation}
where $y_{n}, p_{n}$ are dimensionless, canonically conjugate
coordinates and  momenta
of the $n$th base pair transverse to the chain,
and $V(y) = (1-e^{-y })^{2 }$. $R$ is a dimensionless parameter
which describes the relative strength of on-site vs. elastic interactions; here
$R=10.1$.

The thermodynamic properties of (\ref{eq:trham}) have been
reviewed in Ref.~2.  This work describes the spectra of dynamical
correlations
\begin{equation}
S_{AA} (q, \omega) = \frac{1 }{ N}
\int_{-\infty}^{+\infty}
\frac{dt}{2\pi }  \; e ^ {-i\omega t}
\sum_{m,n} e ^ { iqa(n-m)}
<A_{n}(t)A_{m}(0)>
\label{eq:spectra}
\end{equation}
where, in this paper, mostly $A_{n}=y_{n}$. The integral of
(\ref{eq:spectra}) over all frequencies,
\begin{equation}
S_{AA}(q) \equiv \int_{-\infty  }^{\infty  } d\omega\ S_{AA} (q, \omega)
=\frac{1 }{ N}\sum_{m,n} e ^ { iqa(n-m)}
<A_{n}A_{m}>
\label{eq:staticsq}
\end{equation}
can be computed exactly using the transfer-integral result for the
equal time correlations (cf. Eq. 55 of Ref.~2). It is expedient to
consider normalized spectral functions
\begin{equation}
{\hat S}_{AA} (q, \omega) = \frac{ S_{AA} (q, \omega)}{S_{AA}(q) }
\quad.
\label{eq:nspectra}
\end{equation}
The angular brackets in Eqs. (\ref{eq:spectra})-(\ref{eq:staticsq})
denote canonical ensemble averages. Typically, we implemented
this by repeated molecular dynamics (MD) simulations
of the system  for many different initial conditions, Fourier-transforming
the spatiotemporal correlations obtained from each run, and averaging
over all runs to obtain the final result.

The equations of motion,
\begin{equation}
{\ddot y}_n  = \frac{1 }{ R}\left( y_{n+1}+y_{n-1}-2y_n \right)-V^{'}\left( y_n \right) \quad; \quad n=1,2,... N
\label{eq:eqmot}
\end{equation}
with periodic boundary conditions, $y_0=y_N$, $y_{N+1}=y_1$, and
typical system size $N=1024$,  were numerically integrated
for an interval $T=410$,  using a 4-th order Runge-Kutta algorithm,
with a time step equal to  0.02. Initial conditions were ``canonical", in the sense that (i) the velocities
${\dot y}_n$ were Gaussian-distributed, and (ii) the positions $y_n$
were random variables distributed according to the potential energy
part of the Hamiltonian (\ref{eq:trham}); in addition, the system was
``thermalized" for a certain time, using a Nos{\'e} procedure\cite{dauxpeyr1}.

\section{Spectra: Phonons vs. central peak}
\begin{figure}
\vskip -0truecm a)\hskip5.5truecm b)\hskip-5.5truecm 
\vskip 0truecm\resizebox{\textwidth}{!}
{\includegraphics{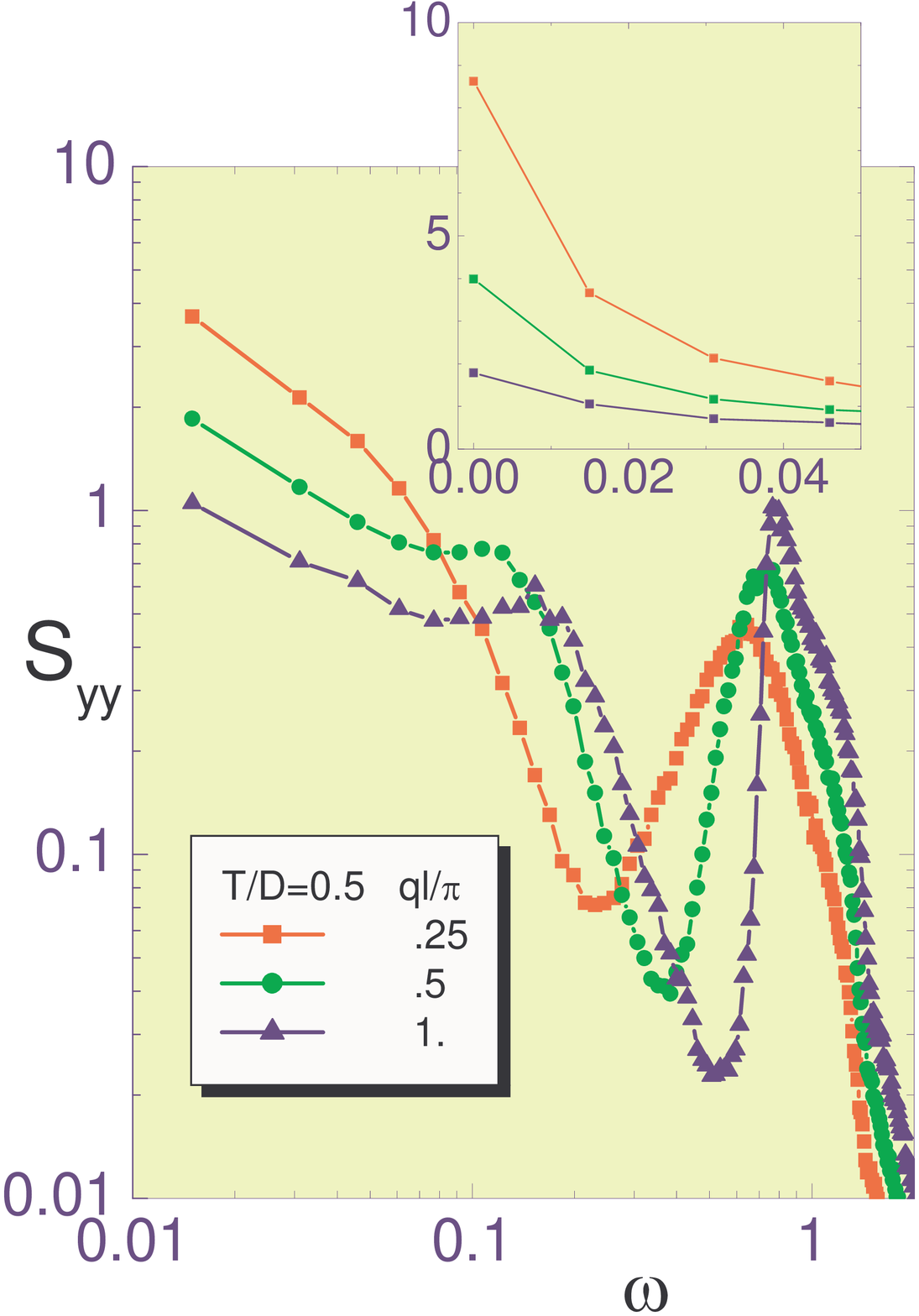}\includegraphics{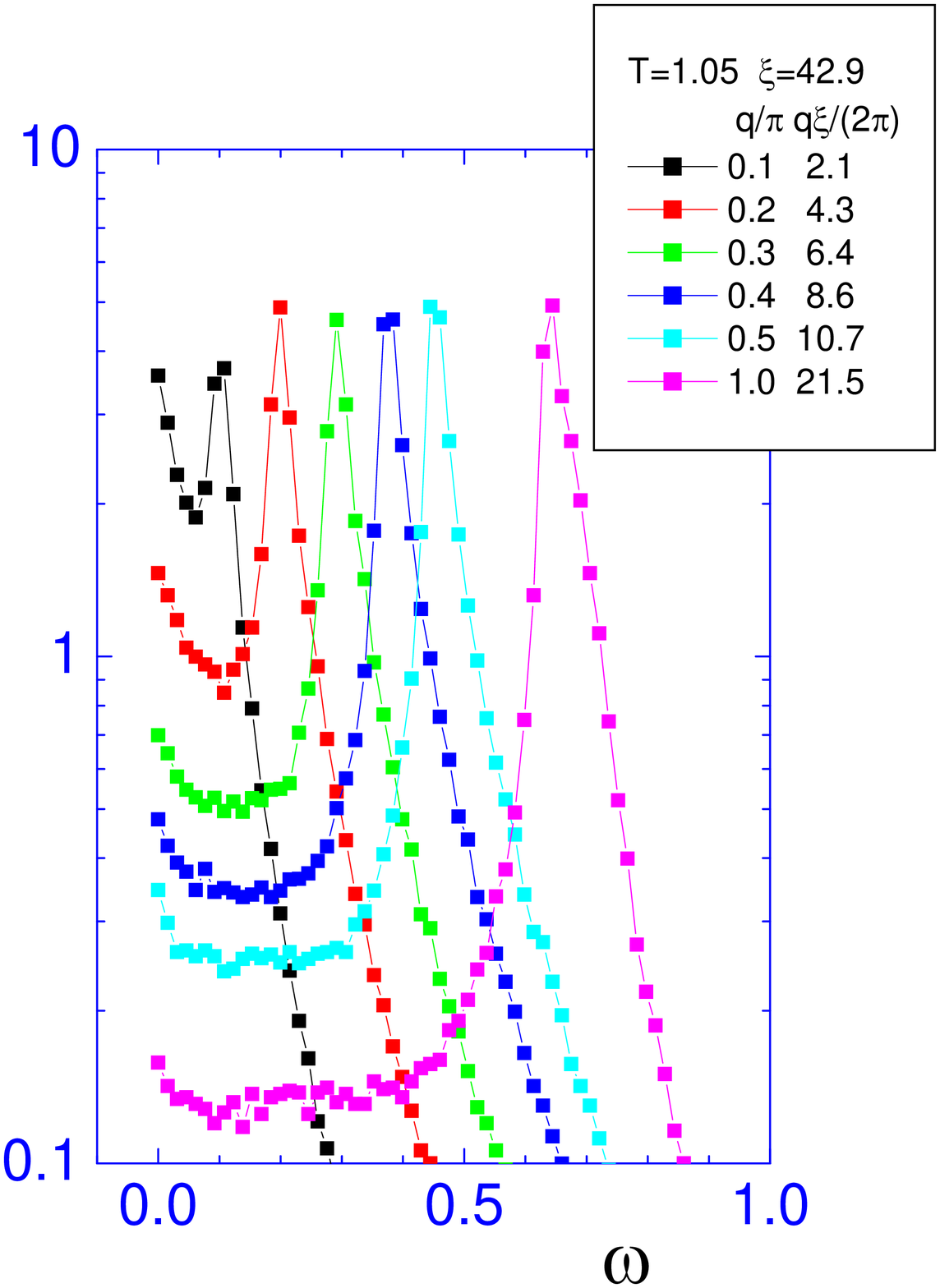}}
\caption { Normalized dynamical correlation spectra. Panel (a)
 presents the results at $T=0.5$, for selected values of
  $q$. The inset shows the zero-frequency details (linear frequency
  scale). Panel (b) presents results at $T=1.05$, for a
  variety of $q$ values.  The inset shows the ratio of the correlation
  length $\xi$ and the wavelength. If the wavelength is smaller than the
  correlation length, the spectrum is dominated by the phonon peak. At
  this length scale, the spectra probe the ``droplets" of the
  high-temperature phase present in the system. Note the gradual
buildup of a central peak at the lowest values of $q$.}
\label{fig:Tp5et95}
\end{figure}

At intermediate temperatures, the spectra are characterized by an
anharmonic phonon component and a strong, low-frequency intensity
(cf. Fig.~\ref{fig:Tp5et95}a); this low-frequency component
becomes even more pronounced at lower values of the wavevector.
There is considerable residual structure in the spectrum; in
particular, a secondary peak at lower frequencies appears to be a
consistent feature.

\begin{figure}
\centerline{\psfig{figure=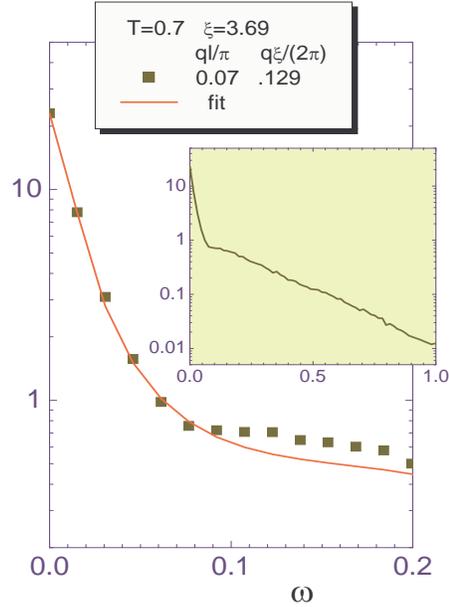,height=8.5truecm,width=6.5truecm}}
\vskip -0.5truecm \caption { Normalized dynamical correlation
spectra at $T=0.7$, and $q/\pi =0.07$.  At long-wavelengths
(compared to $\xi $) the spectrum is dominated by the CP feature.
The inset shows that the tail of the spectra drops off with a
different slope.  The fit has been obtained using
Eq.~(\ref{eq:S1}).} \label{fig:CPTp7}
\end{figure}

As the temperature increases, and the instability approaches, the
structure becomes significantly simpler. The decisive quantity is
the correlation length $\xi$.  If the wavelength is shorter than
$\xi$, i.e. $q\xi /(2\pi )>1$, the spectrum in effect  probes the
"droplets" of the high-temperature phase, of typical size~$\xi$,
which are present in the low-temperature phase; consequently, the 
main feature of Fig.~\ref{fig:Tp5et95}b  is a peak, from the
acoustic phonons\cite{dauxpeyr1}.  At the smallest values of
the ratio $q\xi /(2\pi )$, a central peak (CP) feature begins
to grow, and eventually dominates the spectrum at values 
$q\xi /(2\pi )<<1$; this is the case in Fig.~\ref{fig:CPTp7}.

A first attempt to analyze the data can be made in terms of a
phenomenological relaxation/oscillation spectral function, similar
to the one used in analyzing structural phase transitions\cite{Schwa72}, i.e.
\begin{equation}
{\hat S}_{yy}(q,\omega )=\frac{1 }{\pi \omega  } Im \frac{\omega_{0}^{2 }  }
{\omega_{0}^{2 } - \omega^{2 } -i\omega \Gamma   }
\label{eq:S1}
\end{equation}
where $\Gamma = \Gamma_{0} +   \delta ^{2 }/(\gamma -i\omega )$ is
a relaxational memory kernel, and the $q$-dependence of all the
parameters has been suppressed. If $\Gamma_0\ll\delta^{2}/\gamma$,
it is possible for the spectrum (\ref{eq:S1}) to split into
phonon-like
\begin{equation}
{\hat S}_{yy}(q,\omega ) \approx \frac{1 }{\pi }(1-\rho )
\frac{\omega_{\infty }^{2 }\Gamma_0 } {(\omega^{2 } - \omega_{
\infty }^{2 })^{2 } + (\omega \Gamma_{0})^{2 } } \quad,
\quad
\omega\gg\gamma \label{eq:Sph}
\end{equation}
and CP
\begin{equation}
{\hat S}_{yy}(q,\omega ) \approx \frac{1 }{\pi }\rho
\frac{\gamma'} {  \omega^{2 } + \gamma'^{2 } } \quad,\quad
\omega\ll\gamma \label{eq:SCP}
\end{equation}
contributions, where $\omega_{ \infty }^{2 }=\omega_{0}^{2 }+\delta ^{2 }$,
 $\rho =\delta ^{2 }/\omega_{ \infty }^{2 }$, and
$\gamma'=\gamma (1-\rho )$.

\begin{figure}
  \vskip -0truecm a)\hskip5.5truecm b)\hskip-5.5truecm
  \resizebox{\textwidth}{!}{\includegraphics{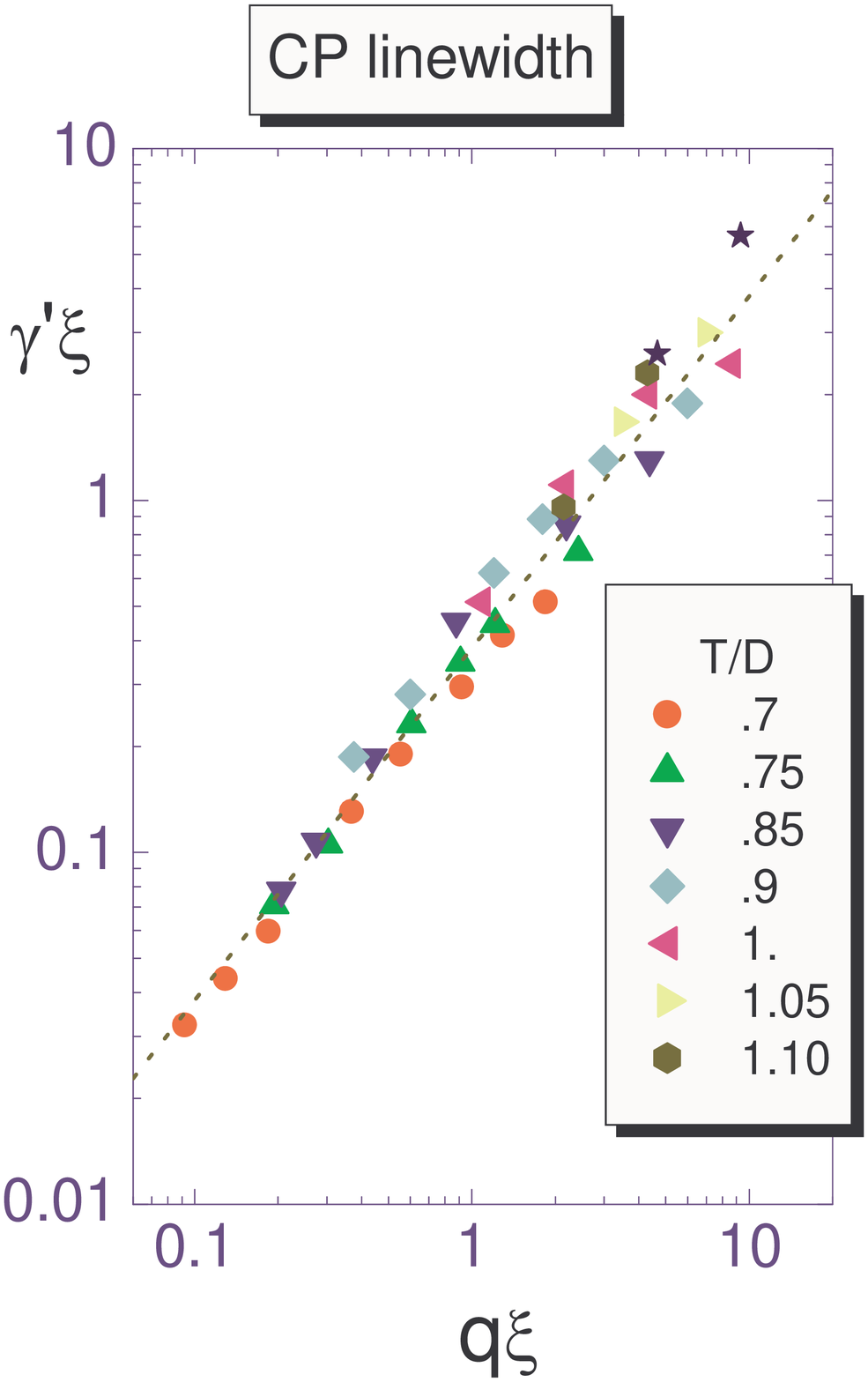}\includegraphics{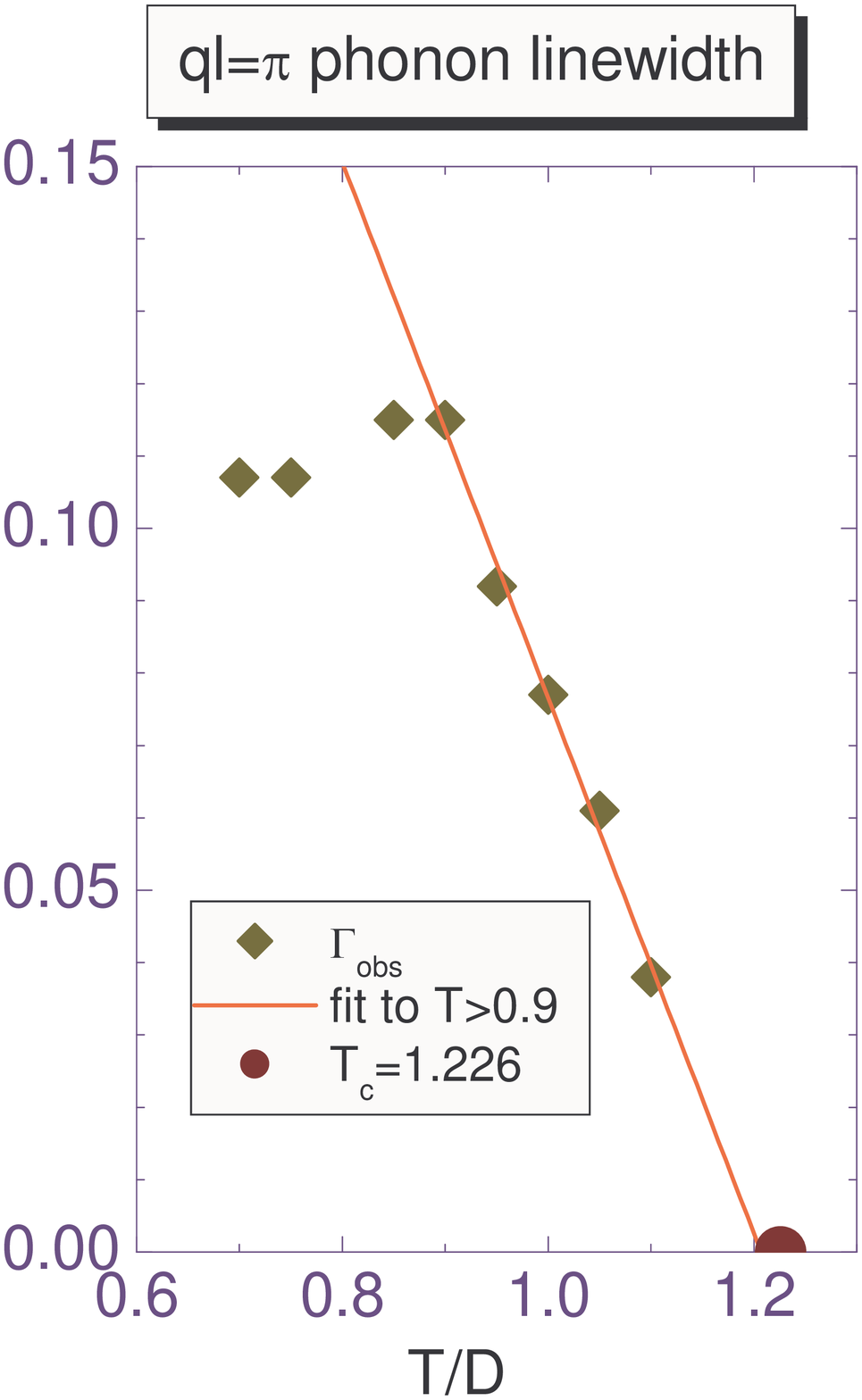}}
\caption{Linewidths. Panel (a) presents the linewidth of the
    central peak as a function of wavelength. The dashed line has unit
    slope. Panel (b) presents the linewidth of the $q=\pi$ phonon as a
    function of temperature. The linewidth extrapolates to zero at a
    temperature not far from $T_{c}$.} \label{fig:width}
\end{figure}

Fig.~\ref{fig:width}a  shows that the CP linewidth $\gamma'$ is
largely independent of temperature; it does however depend on $q$,
roughly linearly, as long as $q\xi/(2\pi)$ does not exceed unity,
i.e. as long as the CP is appreciable.

Returning to the phonon peak, it is possible to follow the
decreasing phonon linewidth as the transition is approached and 
the dynamics - with the exception of the very long wavelengths
$q\xi/(2\pi)\leq 1$
- evolves towards the harmonic limit(Fig.
\ref{fig:width}b). Our data is consistent with a linear critical slowing down, 
i.e.  $\Gamma _{ph} \propto
T_{c}-T$.

\section{Force autocorrelations}
\begin{figure}
\vskip -0truecm a)\hskip5.5truecm b)\hskip-5.5truecm 
\resizebox{\textwidth}{!}{\includegraphics{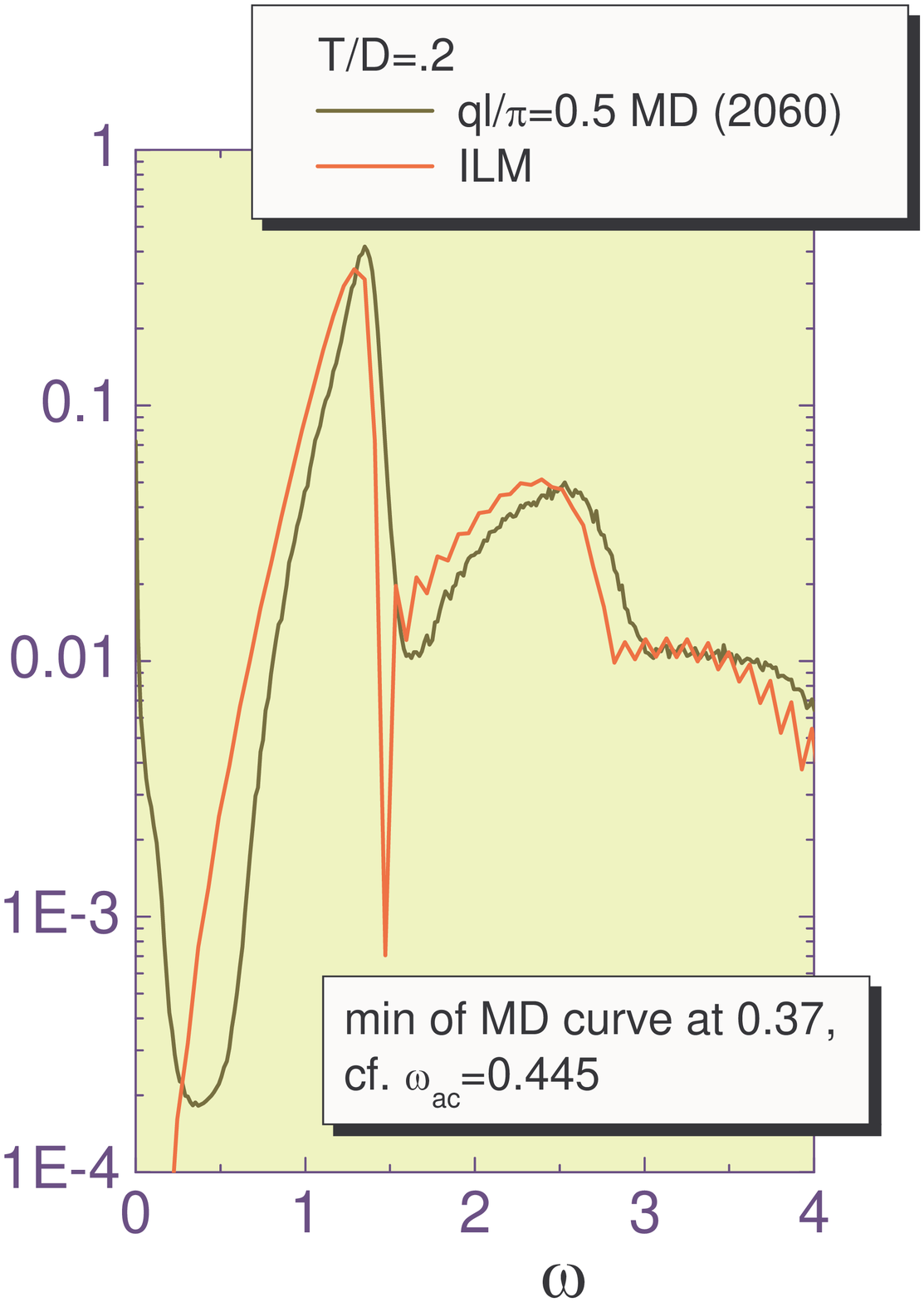}\includegraphics{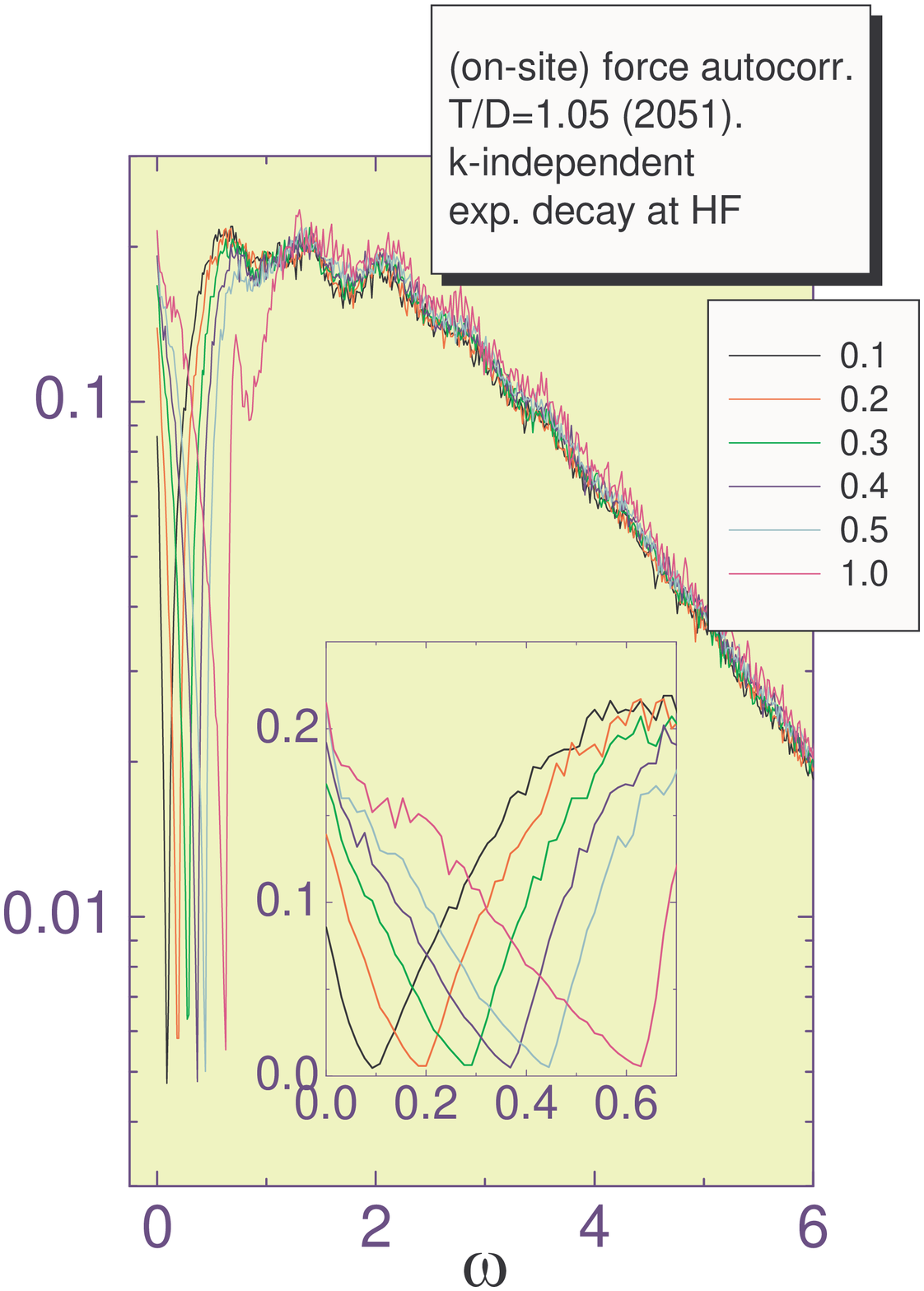}}
\caption {Normalized spectra of force
  autocorrelations. Panel (a) shows the results at low temperatures.
  The wavy curve is a theoretical estimate based on an independent
  localized mode picture, i.e. the normalized spectrum of 
 (\ref{eq:ffILM}).  Panel (b) presents results at higher
  temperatures. Note (i) the extreme dip, almost to zero intensity,
  which occurs almost exactly at the acoustic phonon frequencies
  (inset), and (ii) the exponential decay at higher frequencies. }
\label{spectraforce}
\end{figure}

It is instructive to consider the spectrum of the force
autocorrelations, i.e. $A_n \equiv f_n =-V'(y_n)$ in Eqs.~(\ref{eq:spectra}) -
(\ref{eq:nspectra}). At low temperatures one might try to describe 
the spectra in terms of a crude model of
independent local modes (ILM), i.e. site-independent solutions of the $R \to \infty $
(anticontinuum) limit of (\ref{eq:eqmot}). The explicit form of 
the second time derivative is
\begin{equation}
f_{\lambda ,\delta }(t)=\lambda ^{2 }  \epsilon_{\lambda} ^{1/2 }
\frac{ \cos(\lambda t + \delta ) -  \epsilon_{ \lambda}^{1/2 }  }             
{\left[ 1  -  \epsilon_{\lambda}^{1/2 }
  \cos(\lambda t + \delta ) \right]^{2 }}
\label{eq:ILM}
\end{equation}
where 
$0<\lambda <\lambda _{max}=\sqrt{2}$,  $0<\delta <2\pi $
depend on the initial conditions and $ \epsilon_{\lambda} = 1-\lambda ^{2 }/2$
is the energy of the ILM. Since $R\gg 1$, the above form should
be a good approximation to exact one-site discrete breathers\cite{MarinAubry}. 
The canonical
average of autocorrelations of (\ref{eq:ILM}) is
\begin{equation}
<f(t)f(0)> = \int_{0}^{2\pi } \frac{d\delta}{2\pi }
\int_{ 0}^{\lambda _{max} } d\lambda \> \> Q(\lambda ) 
\> f_{\lambda ,\delta }(t)
f_{\lambda ,\delta }(0)
\label{eq:ffILM}
\end{equation} 
where $Q(\lambda ) = Z^{-1 }\exp\left(-\epsilon_{\lambda} /T\right) $ and 
$Z$ is determined from the normalization condition 
$\int d\lambda \> \> Q(\lambda ) =1$.
The MD force spectra at $T=0.2$ are shown in  Fig.~\ref{spectraforce}a,
along with a numerical Fourier transform of (\ref{eq:ffILM}). Overall
agreement is satisfactory, except for
the very low frequency part of the spectrum.

\begin{figure}
\vskip -1truecm
\centerline{\psfig{figure=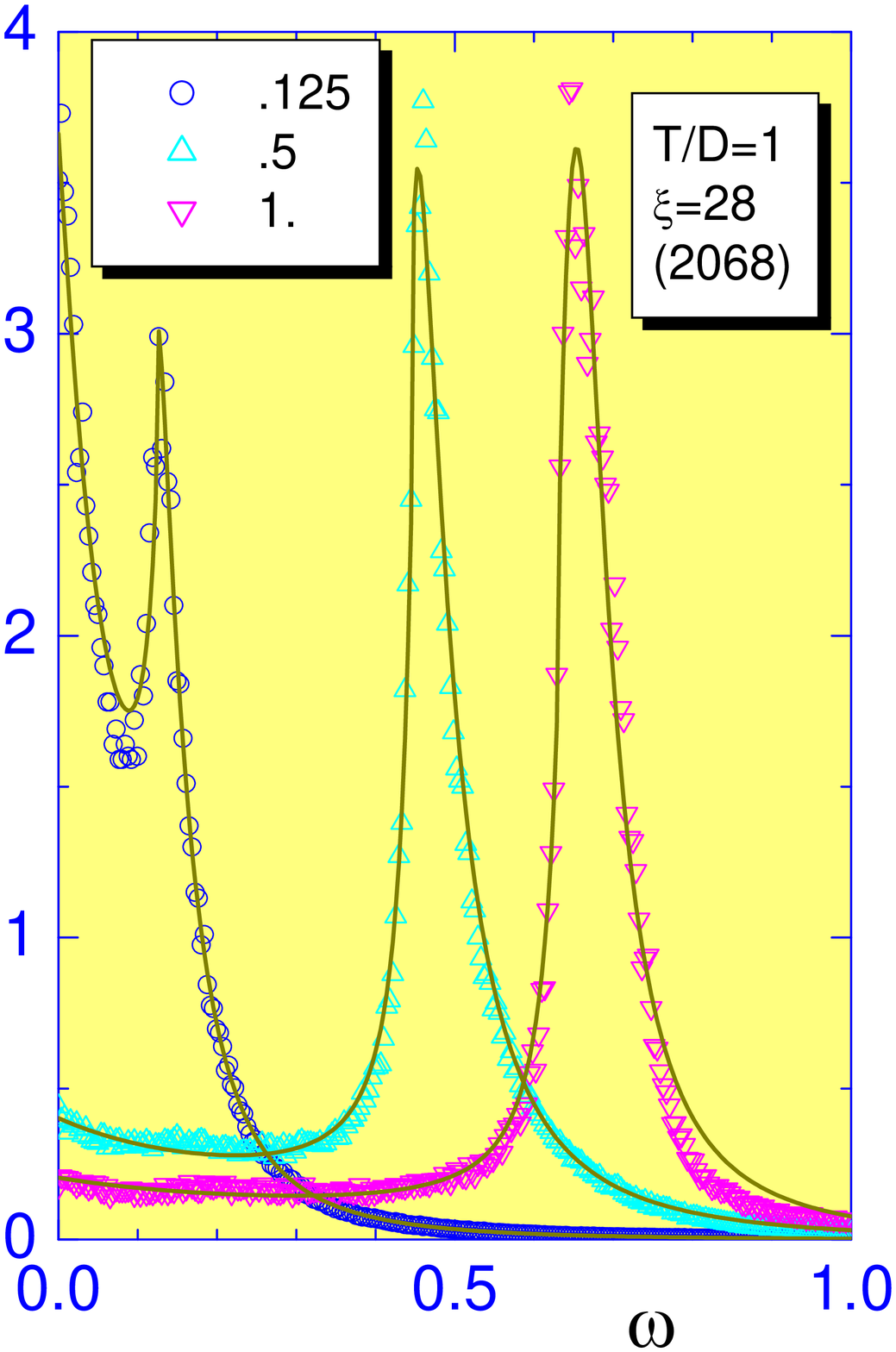,height=7.8truecm,width=6truecm}}
 \vskip -0.5truecm \caption { The spectra $S_{yy}$ at $T=1.0$.
Fits are obtained with the improved Mori-Zwanzig Ansatz, which
incorporates a memory kernel with a dip, similar to the one shown
in Fig.~\ref{spectraforce}b. }
\label{fig:MZfit}
\end{figure}

At higher temperatures, force autocorrelation spectra exhibit the
following features: (i) a very pronounced dip occurring almost
exactly at the bare acoustic frequencies ${\hat
\omega}_q=(2/\sqrt{R}) \sin(q/2)$, characteristic of the
high-temperature phase, and (ii) a roughly $q$-independent decay
at higher frequencies. The observed form of the force spectra
motivates an improved version of  (\ref{eq:S1}), 
with $\omega_{0}^{2 } 
\equiv T/S(q) $, as demanded by
the Mori-Zwanzig projection operator formalism\cite{Forster}, and
$Re \Gamma = \Gamma _{0}\exp(-\omega /\omega _{c}) + \Gamma _{1}$,
where $\Gamma _{1}=a(1-\omega /{\hat \omega} _{q})^{3/2}$ if
$\omega <{\hat \omega} _{q}$, and $\Gamma _1 = b (\omega - {\hat
\omega} _{q})^{1/2} \exp (-\omega /\omega _{c}) $ if $\omega
>{\hat \omega} _{q} $. Preliminary fits obtained with this
improved Ansatz for the memory function (approximating
$Im \Gamma$ by a constant, equal to its value at the peak, 
and setting $\omega _{c}$ equal to the value obtained by
the exponental decay constant of force spectra, cf. above
and Fig. \ref{spectraforce}b)
are shown in Fig.
\ref{fig:MZfit}; they seem to reproduce the MD data much better,
using the same number of adjustable parameters.

\section{Concluding remarks}
The MD data presented show that the critical dynamics of the
Peyrard-Bishop model of DNA thermal denaturation can be thought of
as follows: At length scales shorter than the correlation length
$\xi$, which correspond to ``droplets" of the high temperature
phase, the system reflects the properties of the unstable phase;
oscillatory  dynamics of the soft, acoustic phonons is the result.
The linewidth of these phonons appears to vanish linearly as
$T_c-T$ (``critical slowing  down"). At length scales longer than
the correlation length, the dynamics is dominated by the central
peak.  Fluctuations are stronger, as evidenced from the divergence
of the static structure factor $S(q)$; the typical time scales of
these fluctuations appears however to be non-critical. It remains
a challenge to the theory to establish whether these
``non-critical", $q$-dependent dynamics can be associated with
localized excitations. The preliminary analysis performed at low
temperatures suggests that a picture of independent localized
modes provides a reasonable description of the force
autocorrelations - with the important exception of the very low
frequency regime. Perhaps a more detailed theory of discrete
breather statistical mechanics can improve our understanding of
this part of the spectra.


\end{document}